\documentclass[manuscript,screen,anonymous = FALSE,review = FALSE]{acmart}
\usepackage[normalsize,FIGTOPCAP]{subfigure}

\AtBeginDocument{%
  \providecommand\BibTeX{{%
    \normalfont B\kern-0.5em{\scshape i\kern-0.25em b}\kern-0.8em\TeX}}}

\setcopyright{acmcopyright}
\copyrightyear{2020}
\acmYear{2020}

\acmConference[LAK '12]{Learning Analytics and Knowledge '21}{April 12--16, 2021}{}
\acmBooktitle{Learning Analytics and Knowledge '21,
  April 12--16, 2021}

\begin{document}
\title[Impact of COVID-19-Induced Campus Closure on Self-Regulated Learning]{Measuring the Impact of COVID-19 Induced Campus Closure on Student Self-Regulated Learning in Physics Online Learning Modules}

\author{Tom Zhang}
\email{tom.zhang@ucf.edu}
\affiliation{%
  \institution{University of Central Florida}
  \streetaddress{4111 Libra Dr.}
  \city{Orlando}
  \state{Florida}
  \country{USA}
  \postcode{32816}
}

\author{Michelle Taub}
\email{michelle.taub@ucf.edu}
\affiliation{%
  \institution{University of Central Florida}
  \streetaddress{4111 Libra Dr.}
  \city{Orlando}
  \state{Florida}
  \country{USA}
  \postcode{32816}
}

\author{Zhongzhou Chen}
\email{Zhongzhou.Chen@ucf.edu}
\affiliation{%
  \institution{University of Central Florida}
  \streetaddress{4111 Libra Dr.}
  \city{Orlando}
  \state{Florida}
  \country{USA}
  \postcode{32816}
}

\renewcommand{\shortauthors}{Zhang, et al.}

\begin{abstract}
  This paper examines the impact of COVID-19 induced campus closure on university students' self-regulated learning behavior by analyzing click-stream data collected from student interactions with 70 online learning modules in a university physics course. To do so, we compared the trend of six types of actions related to the three phases of self-regulated learning before and after campus closure and between two semesters. We found that campus closure changed students' planning and goal setting strategies for completing the assignments, but didn't have a detectable impact on the outcome or the time of completion, nor did it change students' self-reflection behavior. The results suggest that most students still manage to complete assignments on time during the pandemic, while the design of online learning modules might have provided the flexibility and support for them to do so.
\end{abstract}



\maketitle

\section{Introduction}
In March of 2020, the majority of higher education institutions across the United States were forced to abruptly close campuses and shift to distance learning for the remainder of the Spring 2020 semester due to the COVID-19 pandemic. As a result, students were suddenly faced with the unusually challenging task of self-regulating their learning activities at home, amidst the disruptions to life brought on by the pandemic. There is widespread concern amongst instructors and administrators regarding the potential negative impact on student learning \cite{Dorn2020, Wilcox2020}, but at the time this manuscript was written, there is little in the way of published literature which quantitatively measures the magnitude or nature of this impact \cite{Gonzalez2020}.

As a result of campus closure, click-stream data from online learning systems has become one of the most reliable sources of data providing information on learning activities and learning outcomes. Several recent studies have analyzed click-stream data to investigate students' self-regulated learning (SRL) processes \cite{Taub2018, Maldonado-Mahauad2018, Li2020}. This current paper introduces our attempt at measuring the impact of COVID-19 induced campus closure on multiple aspects of students' SRL processes, by analyzing click-stream data collected from students enrolled in a university introductory physics class interacting with 70 mastery-based online learning modules (OLMs) as part of the course assignment throughout the Spring 2020 semester.

We will base our data analysis and results interpretation efforts on the theoretical framework of SRL, which models students' SRL processes in three cyclical phases. In the remainder of this section, we will first briefly introduce the SRL framework, present predictions of the impact of campus closure, explain the design of OLMs and OLM sequences, and establish connections between click-stream data from the OLMs and student actions during all three phases of SRL.

\subsection{Frameworks of Self-Regulated Learning}
According to theories of SRL \cite{Zimmerman2013}, a student who is self-regulating is playing an active role in their learning as opposed to being a passive recipient of information. According to Zimmerman's Social Cognitive Theory \cite{Zimmerman2015}, SRL is accomplished by engaging in three cyclical phase during learning: Forethought, Performance, and Self-Reflection. During each of these phases, students use different strategies to monitor and control their learning. The Forethought Phase consists of planning and goal setting, where the student maps out their goals for completing a task and how they are going to achieve them. These decisions are often impacted by students' motivations (e.g., achievement goals). In the Performance Phase, students engage in cognitive learning strategies (e.g., reading content, taking notes) and metacognitive monitoring processes (e.g., time management) to complete tasks. Students are thus enacting their plans and self-monitoring their progress towards those goals. In the Self-Reflection Phase, students evaluate their progress and understanding of the material being studied and assess the factors contributing to their performance (e.g., self-testing). Based on these reflections, students can decide to adapt their behaviors for completing the current or starting subsequent tasks.

These phases are interdependent and thus they do not need to occur in a sequential order, nor do they occur only once during a task. For example, if an online learning module allows multiple attempts at an assessment, a student may choose to adapt how they engage with the content prior to subsequent attempts if self-reflection deemed their initial strategy ineffective. This implies that a student must be \textit{aware} of their own cognition and performance to self-regulate efficiently.

COVID-19 induced campus closure can potentially have multiple negative impacts on a student's SRL processes by both providing fewer opportunities and placing a higher demand for different types of cognitive, metacognitive, and adaptive processes. During the Forethought Phase, a student needs to consider a variety of extraneous factors such as computer access in a family home when planning their study. For the Performance Phase, students face a higher barrier for help seeking \cite{Aleven2016}, while having to more actively monitor the amount of time they spend on each lesson compared to dedicated class hours. In terms of self-reflection, students face lower accessibility for external support such as exchanging notes with classmates or asking questions after class but are still required to evaluate their progress and make adjustments.

\subsection{Measuring SRL from Interactions with Online Mastery-Based Learning Modules (OLMs)}
\subsubsection{Design of OLMs and the OLM sequences}
\begin{figure}[ht]
    \includegraphics[width=.6\textwidth]{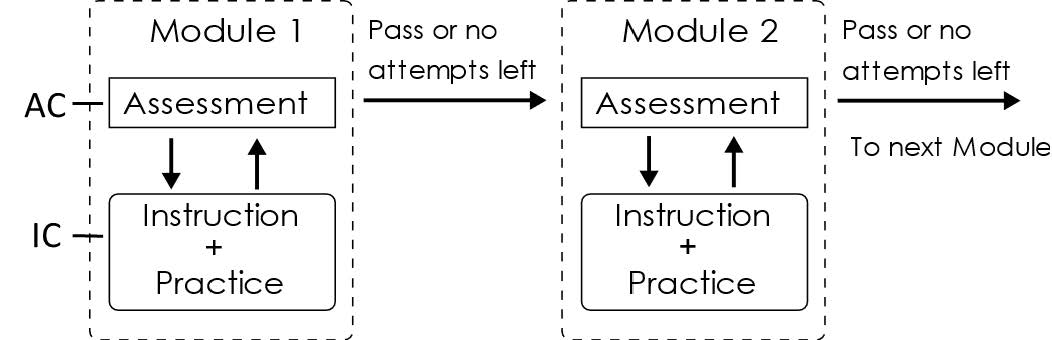}
    \caption{Schematic illustration of OLM design, adapted from \cite{Chen2020}.}
    \label{fig: OLM_design}
    \Description{Design diagram of the Online Mastery-Based Learning Modules.}
\end{figure}

Each OLM is focused on explaining one or two basic concepts, or developing the skills to solve one kind of problem, designed to be completed between 5 to 30 minutes. The OLM consists of an assessment component (AC) which tests students' mastery of the module topic in 1-2 questions, and an instructional component (IC) with instructional text and practice problems on the topic (see Figure \ref{fig: OLM_design}). Upon accessing a module, students are shown the learning objectives of the current module and asked to make an initial attempt on the AC before being allowed to access the IC. Students can make additional attempts on the AC at any time after the first attempt and are not required to access the IC. This design is motivated in part by the "mastery-learning" format \cite{Gutmann, Bloom1968} that allow students who are already familiar with the content to proceed, and by the concept of "preparation for future learning" \cite{Schwartz2005} intending to improve students' learning from the IC by exposing them to the questions first.

A number of OLM modules form an OLM sequence about a more general topic (e.g., conservation of mechanical energy) and students are required to pass the AC or use up all attempts before moving onto the next OLM in that sequence. A typical OLM sequence consists of 5-12 modules and are assigned as self-study homework for students to complete over a period of one to two weeks. In Fall 2019, a total of 44 OLMs were assigned as homework for 7 out of the 10 topics in a calculus-based introductory physics course, while in Spring 2020, 9 out of the 10 topics used a total of 70 OLMs as online homework for the same course. In both semesters, students could earn extra credit by completing some of the OLMs 2-6 days prior to the due date.

\subsubsection{Relating Click-Stream Data to SRL Behavior and Actions}\label{sec: SRL_signal}
Based on the design of the OLMs and OLM sequences, we identify six types of student actions that can be detected or inferred from one or more patterns in click-stream data. These actions are related to or indicative of students' behavior during each of the three phases of SRL, summarized in Table \ref{tab: SRL}.

\begin{table}[ht]
    \centering
    \caption{Relating Data to SRL.}
    \begin{tabular}{p{1.5cm} p{1.55cm} p{4.0cm} p{4.2cm} p{1.35cm}}
        \toprule
        Phase & Behavior & Action/Decision & Data Signal     & Data Level \\
        \midrule
         Forethought & Planning & Skipping, skimming through, or engaging with the initial attempt & Fraction of 1\textsuperscript{st} attempts less than 15\textrm{s}, between 15 and 35\textrm{s}, and longer than 35\textrm{s}. & Module \\
         & Goal Setting & Passing or finishing the OLM with minimal effort. & Fraction of students who adopt a Late Study or No Study strategy. & Module \\
         & & Completing the modules early or close to the sequence due date. & Fraction of OLMs completed at least 1 or 3 days prior to the due date. & Sequence \\
         Performance & Learning & Passing the assessment after studying the module, or passing on a Brief Attempt. & Fraction of students passing after accessing the IC and fraction of Brief Passing Attempts (<35\textrm{s}). & Module \\
         Self-Reflection & Reviewing & Revisiting an upstream module while working within an OLM sequence. & The average number of revisiting events per OLM sequence & Sequence \\
         & & Revisiting a completed OLM before a midterm exam & The number of revisiting events within 3 days of a midterm exam. & Sequence \\
         \bottomrule
    \end{tabular}
    \label{tab: SRL}
\end{table}

Regarding the \textbf{Forethought} Phase, data from the OLMs can provide information on two types of behavior: \textbf{planning} and \textbf{goal setting}. The mandatory first AC attempt on each OLM requires students to plan on whether to engage with the problems or randomly submit a response without reading and proceeding to the IC. Previous studies \cite{Chen2020, Guthrie2020} suggest that attempts submitted under 15 seconds are likely generated by students who skipped reading the problems in the AC, whereas attempts between 15 and 35 seconds are likely generated by students who read the problems but didn't know how to solve them properly. Attempts longer than 35 seconds have a higher probability of being a genuine attempt at solving the AC problems and are more frequently observed among high performing students. The decision to skip the first attempt must be made before the start of the attempt, therefore the fraction of Short (<15\textrm{s}) First Attempts on each OLM provides information on students' \textbf{planning} actions for each OLM.

Furthermore, when a student fails the initial attempt on an OLM, they can decide to study the materials in the IC before attempting the AC again or to make additional attempts immediately. From an SRL perspective, a student with a goal of mastering the content will likely access the IC after 1 or 2 failed attempts on the AC, whereas a student with the goal of completing the module with as little effort as possible are more likely to never access the IC at all (a "No Study" strategy) or access the AC after 3 or more fail attempts (a "Late Study" strategy). Preliminary data analysis suggests that students who are cramming on multiple modules just prior to the due date are more likely to adopt these strategies. Measuring the popularity of Late and No Study strategies among students is used as an indicator for students' \textbf{goal setting} behavior upon accessing each individual OLM.

In addition, students' goal setting action can take place when considering completing an entire OLM sequence as a larger task. In this context, students may set a goal to complete modules early and earn extra credit, or decide to "cram" complete all or most of the modules on or close to the due date. Detailed investigation of students' work distribution as a result of extra credit would require extensive analysis beyond the scope of the current paper (see for example \cite{Felker2020}). In this paper, we present a quick estimation by measuring the number of modules completed at least 1 or 3 days prior to the due date as indicators for students' \textbf{goal setting} behaviors when an OLM sequence is viewed as a task.

Regarding the \textbf{Performance} Phase, it is difficult to infer cognitive strategies adopted by students via click-stream data alone. However, we may straightforwardly estimate the \textbf{outcome of learning} by measuring the percentage of passing AC attempts either before or after accessing the IC. In an OLM sequence, passing attempts before accessing the IC on a later module can be a measure of learning quality of earlier modules \cite{Chen2018}. In a previous study \cite{Chen2020}, we found that some fractions of students pass the AC on a Brief (<35\textrm{s}) Attempt, which could suggest that students guessed the answer by chance or obtained the answer from other sources, such as a classmate. As previously explained, Short (<15\textrm{s}) Attempts are likely generated from students who didn't read the problem body, and Brief Attempts are more likely generated from students who read the problem body. Therefore, we use the fraction of Brief and Short Attempts as an indicator for the \textbf{quality of students' learning} on each OLM.

\textbf{Self-reflection} is mostly a metacognitive process which doesn't often generate direct records in click-stream data. However, we have identified certain types of behaviors which may be indicative of self-reflective processes. Most students interact with each OLM only once and move on after passing the AC, but some will revisit a previously passed module while working on a downstream OLM in the sequence. Therefore, the average number of modules reviewed by one student in a given sequence is chosen as an indicator for the frequency of a self-reflective process. Moreover, self-reflection could take place when students are reviewing for an upcoming exam, and can be estimated by the number of OLMs revisited shortly before an exam day. For the current analysis, we measure the number of modules revisited by a student up to three days before a midterm exam after campus closure as an indicator for \textbf{reviewing} behavior.

It must be emphasized that SRL is an interdependent and iterative process, thus each student action identified is likely influenced by or resulted from multiple different SRL behaviors in different phases. The reason we associate each action to one behavior is just to provide an organizational framework for presenting the results, as well as a baseline for interpreting those results.

\subsection{Examining the Impact of Campus Closure on Student SRL Processes}
In this paper we examine the hypothesis that campus closure resulted in a significant reduction of productive SRL behavior in the student population. More specifically, it would result in the following changes in the six data indicators of SRL actions after campus closure:
\begin{enumerate}
    \item An increase in the frequency of Short 1\textsuperscript{st} Attempts, indicating a reduction in planning and self-assessment.
    \item An increase in the fraction of students adopting a Late or No Study strategy, indicating a shift from mastery-oriented goals to performance-oriented goals.
    \item A decrease in the number of modules completed 3 or more days prior to the due date, indicating fewer students setting goals involving completing the modules early.
    \item A decrease in passing rate before or after studying the IC, or an increase in Short Passing Attempts, indicating a reduction in learning outcomes.
    \item A decrease in the number of revisiting events during each sequence or close to a midterm exam, indicating a decrease in frequency of self-reflection.
\end{enumerate}

We will examine and compare each type of data due before and after campus closure following the data analysis schema explained in the Methods, section (\ref{sec: Methods}). We will also present details about OLMs and their implementation in the physics course as well as operational definitions of actions (e.g., passing a module). We will present the analysis in the Results, section (\ref{sec: Results}), followed by discussion on the implications and possibilities of future studies.

\section{Methods}\label{sec: Methods}
\subsection{OLM Design and Implementation}
The OLM modules are created and hosted on the Obojobo learning objects platform \cite{CenterforDistributedLearning}, an open source online learning platform developed by the Center for Distributed Learning at the University of Central Florida. In the current iteration, the AC of each OLM contains 1-2 multiple choice problems and permits a total of 5 attempts. Each of the first 3 attempts are sets of isomorphic problems assessing the same content knowledge with different surface features or numbers. On the 4\textsuperscript{th} and 5\textsuperscript{th} attempts, students are presented with the same problems in the 1\textsuperscript{st} and 2\textsuperscript{nd} attempts respectively and are awarded 90\% credit. The IC of each module contains a variety of learning resources including text, figures, videos, and practice problems. Access to the IC is locked whenever a student is attempting the AC. Each OLM sequence contains 3-12 OLMs, which students must complete in the order given, with completion defined as either passing the AC or using up all 5 attempts. Readers can access example OLMs in the following URL provided in \cite{obojobo_modules}.

\subsection{Instructional Conditions and Student Population}
In the Spring 2020 semester, 70 OLMs in 9 sequences were assigned as online homework in a calculus-based university introductory physics course, which was taught in a traditional lecture format before campus closure. In Fall 2019, 44 of the 70 modules were assigned in the same course. The new OLMs added in Spring 2020 include sequences S1, S2 (modules 1-16), and modules added to S8 (modules 51-57) and S9 (modules 63-66). Each sequence corresponds to classroom or online instruction for 1-2 weeks with due dates concurrent with lecture instruction. All OLMs in a sequence are due on the same day. In Spring 2020, the last three sequences containing 29 modules are due after campus closure. In Fall 2019, the last 5 modules were due after Thanksgiving break.

%
In Fall 2019, the OLM sequences accounted for 18\% of total course credit, and online homework from a commercial publisher was used for topics for which no OLM module was available. In Spring 2020, the OLM sequences accounted for 36\% of course credit, with no additional homework assignments. In Fall 2019, submissions after the due date received 0 points, while in Spring 2020, late submissions would receive a 13\% daily penalty. In addition, students in both semesters could earn extra credits by completing some OLMs earlier than the due date, as explained in more detail in \cite{Felker2020}.

In Spring 2020, 276 students were initially enrolled in the class consisting of 200 males and 76 females. 107 of the students were historically underrepresented minorities and a total of 263 students passed the course. In Fall 2019, 289 students registered for the course consisting of 234 males and 54 females. 111 of the students were historically underrepresented minorities and a total of 247 students passed the course.

\subsection{Collection and Analysis of Student Log Data}
\subsubsection{Data Collection and Operational Definitions}\label{sec: Definitions}
We list below the operational definition of all key terms related to the data indicators in Table \ref{tab: SRL} and section (\ref{sec: SRL_signal}). Readers interested in more nuanced details of data extraction and cleaning can refer to \cite{Chen2020}.
\begin{itemize}
    \item \textbf{AC Attempt Outcome}: A student passes an AC attempt by answering every question on the AC correctly.
    \item \textbf{AC Attempt Duration}: The time between a student's click on the start attempt button and submission button for a given AC.
    \item \textbf{Brief and Short Attempt}: We will refer to an attempt with duration of less than 15 seconds as a "Short Attempt", and an attempt with duration between 15 and 35 seconds as a "Brief Attempt".
    \item \textbf{Module Pass}: A student will be considered to have passed the module if they passed the AC within 3 attempts. The distinction arises from the fact that the 4\textsuperscript{th} attempt and beyond were already seen by the student and are given reduced credit.
    \item \textbf{Module Fail}: A student will be considered to have failed the module if they fail on all of the first 3 attempts at the AC.
    \item \textbf{Module Complete}: A student either passes the module or uses up all attempts. Time of completion is recorded as the submission time of the first passing attempt or last failed attempt.
    \item \textbf{Late or No Study}: A student does not access the IC before the 3\textsuperscript{rd} attempt of the AC. Students in this category may either access the IC after the 3\textsuperscript{rd} attempt or not, in which case they will be considered to have adopted the "Late Study" or "No Study" strategy respectively. 
    \item \textbf{Module Revisit}: A student interacting with any part of the module for at least 60 seconds after initial completion.
\end{itemize}

\subsection{Analysis Scheme}
From a data analysis perspective, the six types of actions fall into two distinct categories: module level action and sequence level action, as listed in Table \ref{tab: SRL}. Module level actions are actions or decisions made on each module (e.g., planning to skip the first attempt or not). The proportion of module level actions on each module is expected to roughly follow a single linear trend over the semester and to be relatively insensitive to the order of the module in a given sequence. In comparison, sequence level actions are strongly influenced by the location of the module within the sequence (e.g., module completion 3 or more days before the due date) or can only be defined for each sequence (e.g., number of students who revisited at least 2 previous upstream modules in a sequence). The disparity in the number of modules (70) to the number of sequences (9) led us to employ two analysis schemes.

For module level actions, we first calculate the frequency of a given data indicator in each module, then constructed a linear model from each of the two semesters in the form:
\begin{equation}\label{eqn: linear}
    y_i = y_0 +\alpha n_i + \epsilon_i
\end{equation}

where $y_i$ is the frequency of observing the data indicator on module $i$ and $n_i$ the order in which students complete each OLM in the semester. $y_0$ is the intercept, $\alpha$ the slope, and $\epsilon_i$ the noise term which accounts for all other effects not captured by the linear model. When constructing the linear models, the module numbers for Spring 2020 were used for both years as the missing OLM sequences and OLMs were supplemented with online homework assigned from WebAssign.

In addition, data from 2020 was further divided into two segments, the OLMs that were due before and after campus closure, A and B respectively. For comparison, the same partition is applied to the modules in 2019 even though no campus closure took place. Six linear models of the form (\ref{eqn: linear}) were constructed for each of the data indicators outlined in Table \ref{tab: SRL}.

\begin{table}[ht]
    \centering
    \caption{Analysis Acronyms.}
    \begin{tabular}{c c c c}
        \toprule
        Acronym & Type & Year(s) & Segments \\
        \midrule
        20A-B & Within Semester & 2019 & A vs. B \\
        19A-B & Within Semester & 2019 & A vs. B \\
        20A-19A & Between Semesters & 2020 vs. 2019 & A vs. A \\
        20B-19B & Between Semesters & 2020 vs. 2019 & B vs. B \\
        20-19 & Between Semesters & 2020 vs. 2019 & Entire Semester \\
        \bottomrule
    \end{tabular}
    \Description{A table tabulating the analysis acronyms to be used later.}
    \label{tab: Acronym}
\end{table}

Next, we compared the slopes of the linear models for Segments A and B within the same semester listed as 20A-B and 19A-B in rows 1 and 2 of Table \ref{tab: Acronym}. We tested the homogeneity of the regression slopes using Analysis of Covariance (ANCOVA) by including the interaction of due date and module number:
\begin{equation}\label{eqn: ANCOVA1}
    y_i = y_0 + \alpha n_i + \beta\delta_{AB}n_i + \epsilon_i
\end{equation}
where $\delta_{AB} = 0$ if module $i$ is in Segment A, and  $\delta_{AB} = 1$ if the module is in Segment B.  If $\beta$ is not significantly different from zero (i.e., the slope is similar for the two segments), we performed a second ANCOVA of the form:
\begin{equation}\label{eqn: ANCOVA2}
    y_i = y_0 + \alpha n_i + \gamma\delta_{AB} + \epsilon_i
\end{equation}
If $\gamma$ is significantly different from 0, that indicates that the intercepts between the two segments are significantly different. 

If campus closure had a significant impact on a given SRL signal (i.e., either $\beta$ or $\gamma$ is significantly different from $0$), we isolate the effect by comparing the linear models for Segments A and B between the two semesters, listed as 20A-19A and 20B-19B in rows 3 and 4 of Table \ref{tab: Acronym} using the subset of modules common to both semesters. If the effect is directly detectable, we expect that the linear models for Segment B of Spring 2020 to be significantly different from Segment B of Fall 2019.

If no differences were detected for the linear models of Segments A and B, we then proceeded to compare the linear models for the entire semester, row 5 of Table \ref{tab: Acronym}. If the slope or intercept was found to be different, it is likely that either the student population or the instructional condition was different between the two semesters, but campus closure didn't have a detectable impact on the action analyzed.

Analysis of data on sequence-level actions (i.e., early completion and revisiting) is more straightforward. We first record the observation of the data indicator (e.g., completing a module 3 days before the due date) for every student who accessed all modules in a given sequence. The Friedman test is then performed to observe any differences between the sequences over each semester. Each sequence can be treated as an independent category since they cover different topics and have different due dates. To satisfy the complete block design requirement of the Friedman test, only students who accessed all 9 sequences were retained in the analysis. In the case of revisiting, where we count the number of students which have at least one revisiting event, Cochran's Q test was used in lieu of the Friedman test.

If statistically significant differences are detected between sequences, a post hoc analysis using pairwise exact tests \cite{Eisinga2017} (early completion) or McNemar's tests (revisiting) is conducted to determine the precise differences in frequencies between the sequences. If campus closure had a significant impact on students' SRL behavior related to the observed actions, then the observed frequency on sequences due after campus closure will be significantly different from those due before campus closure. Data from Fall 2019 was also analyzed and presented for revisiting actions but not for early completion actions since the first 5 or 6 modules of S8 and S9 were not available, inevitably resulting in significantly less early completion events in those sequences.

For the action of revisiting before an exam, we simply recorded the number of modules revisited over a period of three days leading to an exam by each student and compared the distribution between the semesters via Wilcoxon test. All statistical procedures were conducted in R \cite{RCoreTeam2019} with the tidyverse and PMCMRplus packages \cite{tidyverse, PMCMRplus}.

\section{Results}\label{sec: Results}
For each figure in this section, the black vertical line separates the modules due before campus closure (Segment A), from those due after (Segment B). For figures representing module level data, the blue line visualizes the linear regression models with the shaded areas representing the 95\% confidence interval. For each table in this section, statistically significant differences are emboldened and appended with asterisk markings \textbf{*}, \textbf{**}, \textbf{***} representing significance at the $\alpha = 0.05,\;0.01,\;\text{and}\;0.001$ levels respectively. 

\subsection{Overall Engagement}
\begin{figure}[ht]
    \centering
    \subfigure[Spring 2020]{
        \includegraphics[width=.43\textwidth]{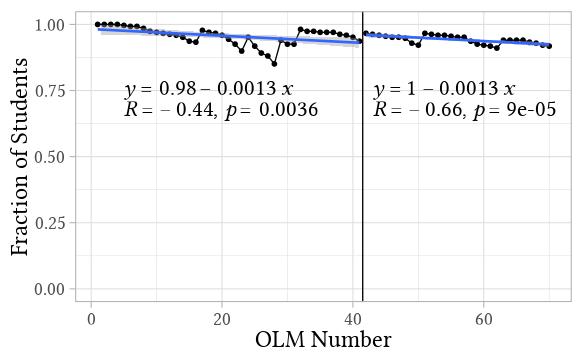}
    }
    \subfigure[Fall 2019]{
        \includegraphics[width=.43\textwidth]{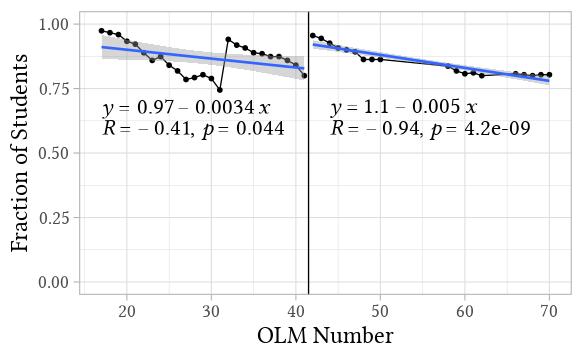}
    }
    \caption{Fraction of students accessing each module.}
    \Description{Two graphs displaying the fraction of students who access each module, one for Spring 2020 and one for Fall 2019.}
    \label{fig: Engagement}
\end{figure}

The fraction of student module access is shown in Figure \ref{fig: Engagement} with the linear models constructed for Segments A and B for each semester. The ANCOVA of the linear models following the analysis scheme outlined in Table \ref{tab: Acronym} are shown in Table \ref{tab: Engagement}.

\begin{table}[ht]
    \centering
    \caption{ANCOVA statistics for overall engagement.}
    \begin{tabular}{c c c c c}
        \toprule
        & 20A-B & 19A-B & 20A-19A & 20B-19B \\
        \midrule
        F (slope) & 0.003 & 0.895 & \textbf{4.996*} & \textbf{48.339**} \\
        F (intercept) & \textbf{7.073**} & \textbf{14.416**} & N/A & N/A \\
        \bottomrule
    \end{tabular}
    \Description{A table outlining the ANCOVA statistics for engagement rates.}
    \label{tab: Engagement}
\end{table}

The fraction of students accessing each module remained remarkably stable over the entire 2020 semester, between 85\% and 100\% with no significant difference detected in the slopes of the linear models. The linear model for Segment B (post-campus closure) has a higher intercept than that of Segment A (pre-campus closure), possibly due to modules 29 and 30 having lower than average access percentage. Similarly, data from Fall 2019 had no significant difference in the slopes of the linear models between each segment, but had a higher intercept in the latter half. The regression slopes in 2019 were significantly more negative than their 2020 counterparts, despite the absence of campus closure

\subsection{The Forethought Phase}
\subsubsection{Planning}
\begin{figure}[ht]
    \centering
    \subfigure[Spring 2020]{
        \includegraphics[width=.43\textwidth]{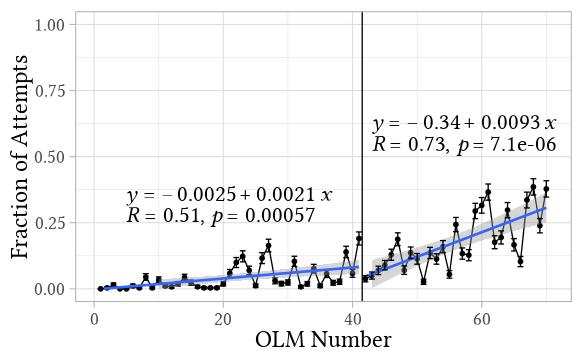}
    }
    \subfigure[Fall 2019]{
        \includegraphics[width=.43\textwidth]{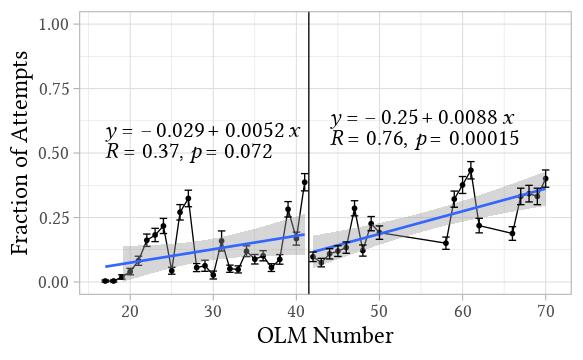}
    }
    \caption{Fraction of first attempts under 15 seconds.}
    \Description{Two graphs displaying the fraction of first attempts which are under 15 seconds for each module, one for each semester.}
    \label{fig: First15}
\end{figure}

In Figure \ref{fig: First15}, we plot the fraction of 1\textsuperscript{st} AC attempts on each module under 15 seconds as an indicator for students' planning action before each OLM. The results of the comparisons between linear models are listed in Table \ref{tab: First15}.

\begin{table}[ht]
    \centering
    \caption{ANCOVA statistics for Short First Attempts.}
    \begin{tabular}{c c c c c}
        \toprule
        & 20A-B & 19A-B & 20A-19A & 20B-19B  \\
        \midrule
        F (slope) & \textbf{23.034***} & 1.168 & 1.032 & 0.066  \\
        F (intercept) & N/A & 2.418 & \textbf{7.809**} & 3.297 \\
        \bottomrule
    \end{tabular}
    \Description{A table outlining the ANCOVA statistics for short first attempts.}
    \label{tab: First15}
\end{table}

In 2020, the proportion of Short First Attempts increased significantly more rapidly in Segment B than in Segment A. This rapid shift in slope is not detected in data from Fall 2019, for which there was no significant difference observed between the intercepts of the regression lines in each segment. Our analysis also failed to detect significant differences between the linear models for Segment B between the two semesters.

\begin{figure}[ht]
    \centering
    \subfigure[Spring 2020]{
        \includegraphics[width=.43\textwidth]{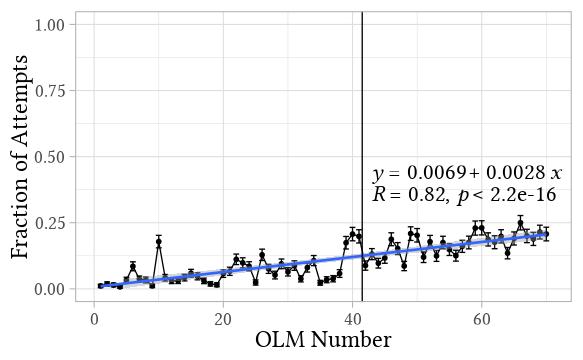}
    }
    \subfigure[Fall 2019]{
        \includegraphics[width=.43\textwidth]{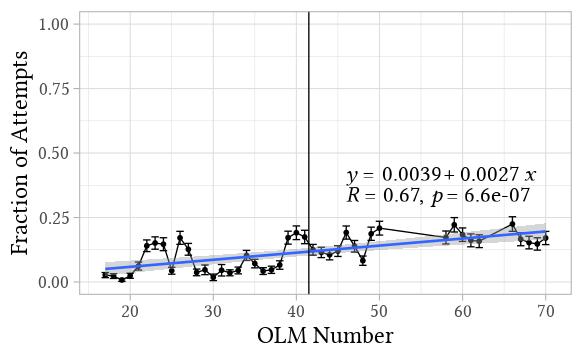}
    }
    \caption{Fraction of first attempts between 15 and 35 seconds.}
    \Description{Two graphs, one for 2020 and one for 2019, displaying the fraction of first attempts between 15 and 35 seconds per module.}
    \label{fig: First35}
\end{figure}

In contrast, the fraction of 1\textsuperscript{st} attempts between 15 and 35 seconds showed no difference in either the slopes or intercepts between either Segments A and B of each semester, or the overall linear models for both semesters, see Figure \ref{fig: First35}.

\subsubsection{Goal Setting}
\begin{figure}[ht]
    \centering
    \subfigure[Spring 2020]{
        \includegraphics[width=.43\textwidth]{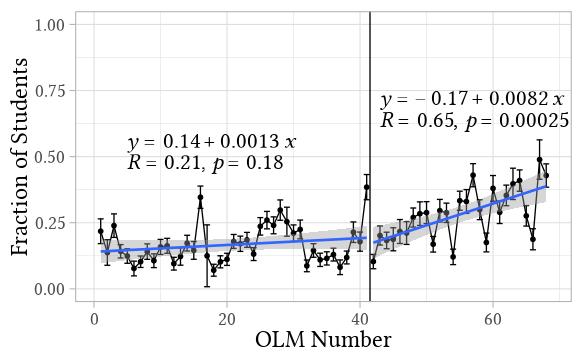}
    }
    \subfigure[Fall 2019]{
        \includegraphics[width=.43\textwidth]{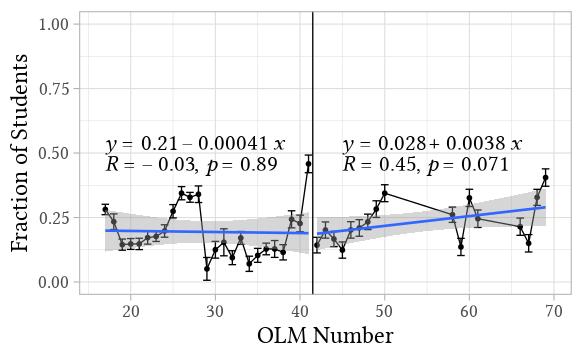}
    }
    \caption{Fraction of students adopting a Late Study or No Study strategy.}
    \Description{Two graphs displaying the fraction of students who adopt a Late Study or No Study strategy per module, one for each semester.}
    \label{fig: LNS}
\end{figure}

In Figure \ref{fig: LNS}, we plot the fraction of students who adopted either a Late or No Study strategy on each module as an indicator for students' goal setting behavior for each module.

In Spring 2020, the number of students adopting a Late or No Study strategy increased much faster in Segment B when compared to Segment A, $F_{1,67}=10.617,\;p<0.05$, while no difference in trend was detected for Fall 2019. Comparing the linear models for each segment between semesters did not show a statistically significant difference in either the slopes or the intercepts of each model.

\begin{figure}[ht]
    \centering
    \subfigure[Spring 2020 (1 Day)]{
        \includegraphics[width=.43\textwidth]{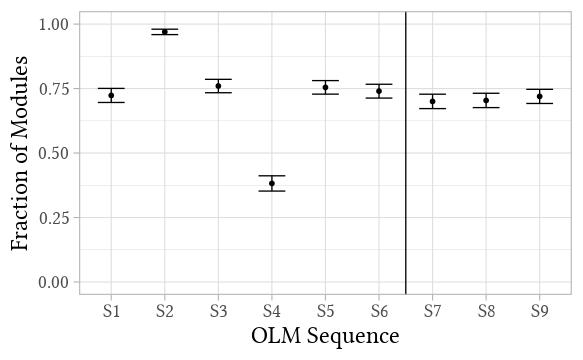}
    }
    \subfigure[Spring 2020 (3 Days)]{
        \includegraphics[width=.43\textwidth]{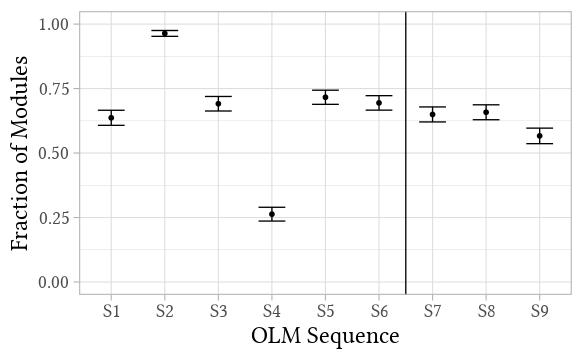}
    }
    \caption{Average fraction of modules completed in a sequence at least 1 or 3 days before the due date.}
    \Description{Two graphs, one displaying the fraction of modules completed 1 day early per sequence in 2020, one displaying the fraction of modules completed 3 days early per sequence in 2019.}
    \label{fig: Early}
\end{figure}

To examine students' sequence level goals, we plot the average number of modules completed by a student at least 1 or 3 days before the sequence due date, see Figure \ref{fig: Early}. Friedman's test detected that there were significant differences in the fractions between different sequences in both conditions ($\chi^2_{1}(8)=446.121,\;p<0.001$ and $\chi^2_{3}(8)=586.571,\;p<0.001$), but post hoc analysis showed that none of the significantly different sequences were due after campus closure.

\subsection{The Performance Phase}
\subsubsection{Learning Outcome}
\begin{figure}[ht]
    \centering
    \subfigure[Spring 2020]{
        \includegraphics[width=.43\textwidth]{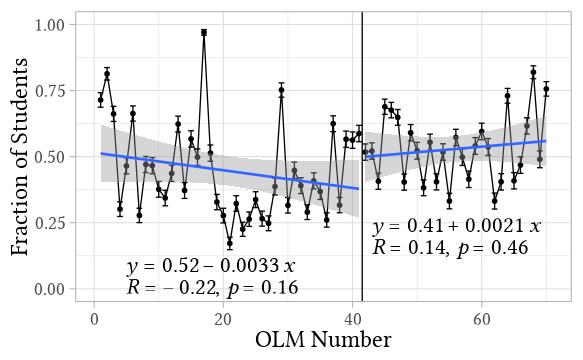}
    }
    \subfigure[Fall 2019]{
        \includegraphics[width=.43\textwidth]{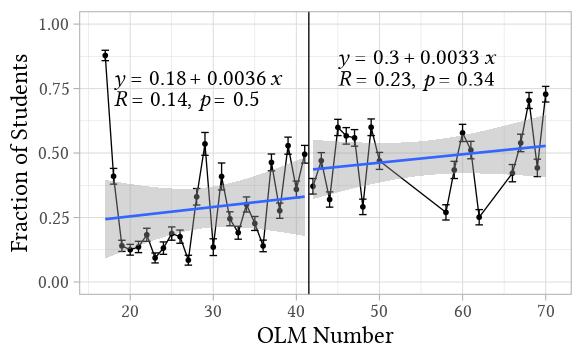}
    }
    \caption{Fraction of students who passed before study of the IC.}
    \Description{Two graphs displaying the fraction of students passed the module before studying the instructional components, one for 2020 and one for 2019.}
    \label{fig: BSP}
\end{figure}

The fraction of students passing each module before accessing the IC, as plotted in Figure \ref{fig: BSP}, remained largely stable throughout Spring 2020. No significant difference was found between the regression slopes for Segments A and B. However, the intercept of Segment A was found to be higher than that of Segment B, $F_{1,67}=4.153,\;p<0.05$, likely caused by the slightly negative slope of the model in Segment A. In 2019, both the regression slopes and intercepts remain not significantly different between both segments. Comparing Segment A between semesters, we found no significant difference for the regression slopes, but the intercept in Spring 2020 was higher than that of Fall 2019, $F_{1,47}=5.225,\;p<0.05$. No difference was detected in Segment B between each semester.

\begin{figure}[ht]
    \centering
    \subfigure[Spring 2020]{
        \includegraphics[width=.43\textwidth]{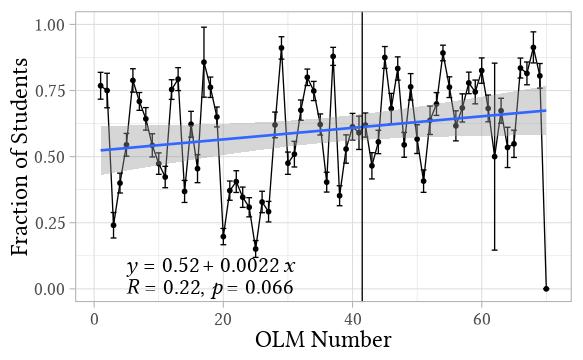}
    }
    \subfigure[Fall 2019]{
        \includegraphics[width=.43\textwidth]{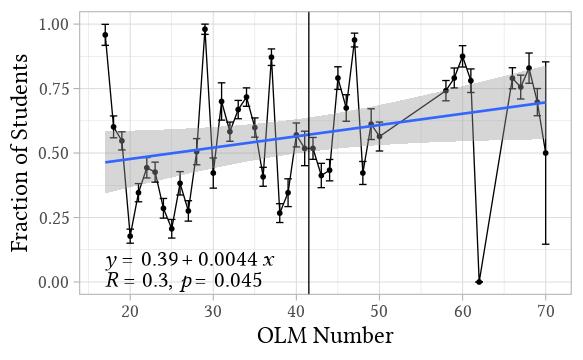}
    }
    \caption{Fraction of students who passed after study of the IC.}
    \Description{Two graphs displaying the fraction of students passed the modules after studying the instructional components, one for 2020 and one for 2019.}
    \label{fig: ASP}
\end{figure}

The fraction of students passing each module after accessing the IC is remarkably stable across both semesters, with no difference detected in all of our comparisons (see Figure \ref{fig: ASP}).

\begin{figure}[ht]
    \centering
    \subfigure[Spring 2020]{
        \includegraphics[width=.43\textwidth]{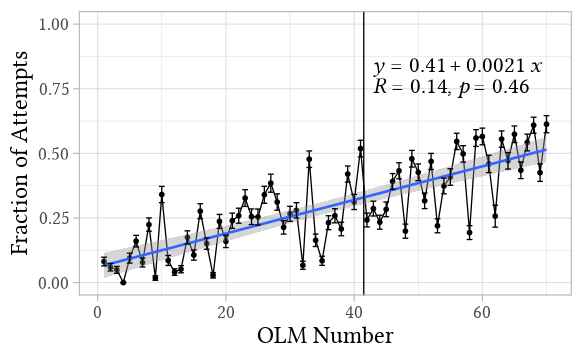}
    }
    \subfigure[Fall 2019]{
        \includegraphics[width=.43\textwidth]{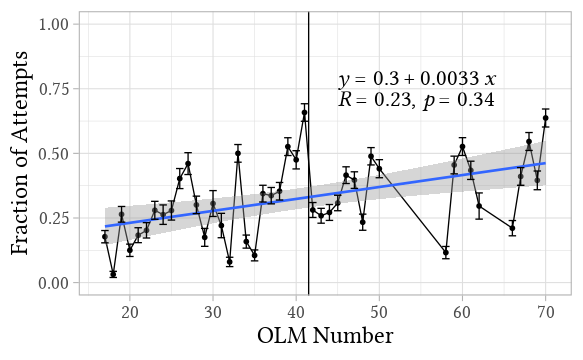}
    }
    \caption{Fraction of Brief Passing Attempts.}
    \label{fig: Brief}
    \Description{Two graphs displaying the fraction of passing attempts which are between 15 and 35 seconds, one for 2020 and one for 2019.}
\end{figure}

In contrast, the percentage of Short Passing Attempts saw a significant upward shift in Segment B of Spring 2020,$F_{1,66}=10.667,\;p<0.05$, but not in Fall 2019. No difference was found when comparing the linear models for 20B-19B. No difference in intercepts was found for comparisons 20A-19A, and 20B-19B. When the criterion is loosened to include Brief Attempts, no difference in slope nor intercept was found between the models, see Figure \ref{fig: Brief}.

\begin{figure}[ht]
    \centering
    \subfigure[Spring 2020]{
        \includegraphics[width=.43\textwidth]{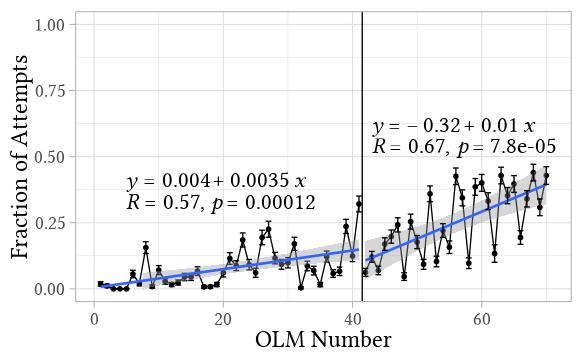}
    }
    \subfigure[Fall 2019]{
        \includegraphics[width=.43\textwidth]{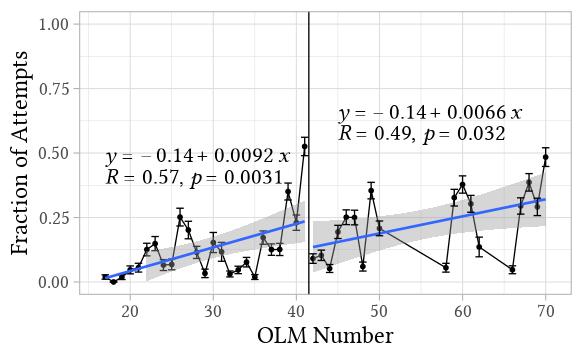}
    }
    \caption{Fraction of Short Passing Attempts.}
    \Description{Two graphs displaying the fraction of passing attempts which are under 15 seconds, one for 2020 and one for 2019.}
    \label{fig: Short}
\end{figure}

\subsection{The Self-Reflection Phase}
\begin{figure}[ht]
    \centering
    \subfigure[Spring 2020]{
        \includegraphics[width=.43\textwidth]{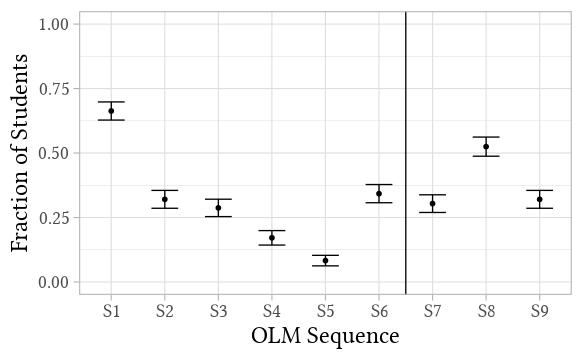}
    }
    \subfigure[Fall 2019]{
        \includegraphics[width=.43\textwidth]{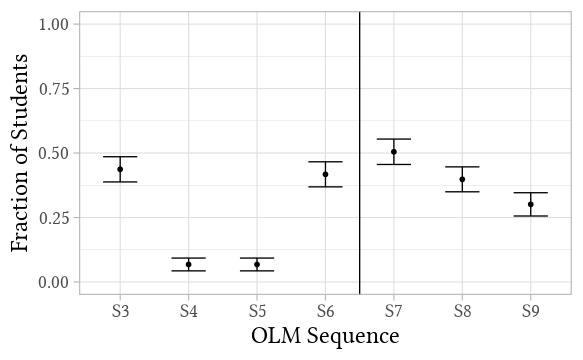}
    }
    \caption{Fraction of students with at least 1 revisiting event.}
    \Description{Two graphs displaying the fraction students with at least one revisiting event by sequence, one for 2020 and one for 2019.}
    \label{fig: RevisitSeq}
\end{figure}

In Figure \ref{fig: RevisitSeq}, we plot the fraction of students with at least 1 revisiting event as defined in section (\ref{sec: Definitions}). Cochran's Q test indicated that there were significant differences between the OLM sequences in each semester, $\chi^2_{2020}(8)=222.514,\;p<0.001$ and $\chi^2_{2019}(8)=92.888,\;p<0.001$. Pairwise McNemar tests determined that in 2020, the revisiting fraction of S1 and S8 was significantly higher than the rest of the sequences and that of S5 was lower than all the other sequences except S4. Additionally, it was found in 2019, S4 and S5 were significantly lower than the rest of the sequences.

\begin{figure}[ht]
    \centering
    \subfigure[Spring 2020]{
        \includegraphics[width=.43\textwidth]{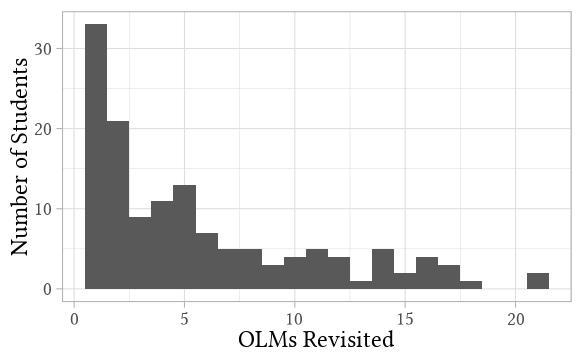}
    }
    \subfigure[Fall 2019]{
        \includegraphics[width=.43\textwidth]{figures/Results/4. Self-Reflection/exam2_20.jpg}
    }
    \caption{Histograms of the number of modules revisited up to three days before the second midterm exam.}
    \Description{Two histograms, one for 2020 and one for 2019, which display the distribution of students among modules revisited.}
    \label{fig: RevisitExam}
\end{figure}

The distribution of modules revisited by each student within a 3 day period leading to the second midterm exam is plotted in Figure \ref{fig: RevisitExam}. There does not appear to be a statistically significant difference in the distribution of the number of modules revisited by students within 3 days of the exam according to the Wilcoxon Test (Z=8115.5, p=0.696).

\section{Discussion}\label{sec: Discussion}
\subsection{Discussion of Results}
Results of the current analysis indicate that some SRL actions are impacted much more by COVID-19 induced campus closure than others; overall, the changes in SRL action that can be attributed to it were less than expected.

Most notably, we saw a significant increase in trend of the fraction of Late or No Study events per module for the OLMs due after campus closure, which was not observed in Fall 2019. Similarly, an increasing trend was observed for the percentage of Short Attempts, indicative of guessing or copying, for the OLMs due after campus closure which was again absent in 2019. These observations suggest that after campus closure, more students are adopting performance-oriented goals (e.g. completing the module with as little time as possible) over mastery-oriented goals (e.g. internalizing new content) in the Forethought Phase and executing those strategies during the Performance Phase. Notably, there were no sudden shifts in the trend of Brief Attempt submissions, which are more likely to be generated from students who read the assessment problem before deciding to guess. Therefore, the abrupt increase in Short Attempts is more likely a result of change in student strategies, rather than an increase in content difficulty.

It must be mentioned that in both cases we did not find a significant difference comparing data from Segment B between the 2020 and 2019 semesters. It could mean that the impact from campus closure was not strong enough to be detected, but could also be caused by a lack of statistical power of the analysis resulting from the smaller number of modules released in Segment B of 2019.

A similar increase in slope after campus closure was observed for the 1\textsuperscript{st} attempts under 15 seconds for each OLM, which is indicative of students' planning action during the Forethought Phase. This could suggest that more students planned to skip the self-assessment opportunities before starting each new module. However, a similar trend was found for the same modules in Fall 2019 albeit to a lesser extent. This may imply that the observed change could in part be due to increase in content difficulty or reduction in engagement towards the end of the semester unrelated to campus closure.

We found no significant differences comparing a number of data indicators before and after campus closure: the number of modules completed earlier than the due date, module passing rate before and after study, the number of module revisiting events. One exception found was the fraction of students revisiting sequence S8, which seems to be higher than average in Spring 2020 which is opposite to the predicted impact of campus closure.

In summary, our results show that COVID-19 related campus closure and distance learning affected how students completed or planned to complete each module, pushing more students towards adopting goals and strategies that minimize time and effort required to pass each module. On the other hand, we found that there was little or no impact on overall engagement, student performance, or self-reflective processes.

It is worth noting that the fraction of students accessing each module decreased more rapidly in both Segment A and B in Fall 2019 when compared to Spring 2020. Additionally, the fluctuation in module access is much greater in 2019. This difference cannot be caused by the 2020 campus closure and are more likely due to differences in instructional conditions and student populations between the semesters, which could have a non-negligible impact on our analysis.

\subsection{Implication of Results}
Despite the significant disruptions caused by the COVID-19 induced campus closure, the impact on students' SRL processes in online learning is less than what many have feared. Even though an increasing number of students adjusted their plans and strategies towards preserving time and resources after campus closure, the fraction of students completing and passing the OLMs on time or early remained remarkably stable, as well as the fraction of students who revisited a previously passed module.

This could imply that college students enrolled in a first year physics course have stronger self-regulatory skills than we previously thought. It could also suggest that online learning may have provided students with the needed flexibility to adjust to unexpected disruptions, which is consistent with findings in \cite{Gonzalez2020}. Furthermore, it is possible that the mastery-based design of the OLM sequences could have facilitated students' SRL processes by providing frequent self-assessment opportunities during learning. Of course, many other factors such as difference in instructional condition and student populations could also have contributed to the lack of observed differences.

\subsection{Limitations and Future Directions}
The current paper provides a quick estimation of the impact of campus closure on students' SRL behavior in an online learning environment. Many of our decisions regarding data selection and analysis methods were prioritized for quickly detecting trends in the data rather than creating a comprehensive model. Those choices, while sufficient for the purposes of the current study, leave much room for improvement in future studies.

First, future studies could include more data such as duration of study and number of practice problems. While containing rich information about students' learning, they are not included in this paper as their relation with quality of learning is less straightforward. Furthermore, incorporation of data from multiple sources such as the Revised Achievement Goal Questionnaire-Revised \cite{Elliot2008} could be used to measure aspects of students' SRL (i.e., motivation) that are not well reflected in click-stream data.

Second, future studies should extend beyond the current linear regression models used to quickly estimate shifts in data. More sophisticated models such as linear mixture modeling or the ones used in \cite{Chen2020} could account for a number of factors overlooked in the current study. Such factors include the difference in topical difficulty between different OLMs and OLMs sequences or differences in instructional policy choices in courses. Additionally, individual students' shifts in study strategies could be tracked for analysis on a finer scale.

Third, the validity of comparisons of data between Fall 2019 and Spring 2020 is less than ideal, in large part due to differences in the number of modules assigned. Future studies involving the latest data from Fall 2020, during which the entire course was taught online, could provide a better baseline for comparison.

Finally, the current study paper presents a case study that includes students from one class, studying one topic, using one type of online instructional design. A highly valuable direction of future research is to compare and contrast multiple studies involving different student populations, subject matter, and online instructional designs to obtain generalizeable  knowledge that will guide the design of future learning environments. Directed by new insights from such analysis, these systems can not only be more resilient to disruptions, but also more flexible in accommodating today's increasingly diverse student population \cite{U.SDepartmentofEducation2017}.

\begin{acks}
This work is supported by NSF Award No. DUE-1845436. We would like to thank UCF Center for Distributed Learning for creating the UCF Open project and the Obojobo platform, in particular Dr. Francisca Yonekura, Ian Turgeon and Zachary Berry. 
\end{acks}
\bibliographystyle{ACM-Reference-Format}
\bibliography{bibliography}


\begin{thebibliography}{24}


\ifx \showCODEN    \undefined \def \showCODEN     #1{\unskip}     \fi
\ifx \showDOI      \undefined \def \showDOI       #1{#1}\fi
\ifx \showISBNx    \undefined \def \showISBNx     #1{\unskip}     \fi
\ifx \showISBNxiii \undefined \def \showISBNxiii  #1{\unskip}     \fi
\ifx \showISSN     \undefined \def \showISSN      #1{\unskip}     \fi
\ifx \showLCCN     \undefined \def \showLCCN      #1{\unskip}     \fi
\ifx \shownote     \undefined \def \shownote      #1{#1}          \fi
\ifx \showarticletitle \undefined \def \showarticletitle #1{#1}   \fi
\ifx \showURL      \undefined \def \showURL       {\relax}        \fi
\providecommand\bibfield[2]{#2}
\providecommand\bibinfo[2]{#2}
\providecommand\natexlab[1]{#1}
\providecommand\showeprint[2][]{arXiv:#2}

\bibitem[\protect\citeauthoryear{??}{obo}{[n.d.]}]%
        {obojobo_modules}
 \bibinfo{year}{[n.d.]}\natexlab{}.
\newblock
\newblock
\urldef\tempurl%
\url{https://canvas.instructure.com/courses/1726856}
\showURL{%
\tempurl}


\bibitem[\protect\citeauthoryear{Aleven, Sewall, Popescu, Ringenberg, van
  Velsen, and Demi}{Aleven et~al\mbox{.}}{2016}]%
        {Aleven2016}
\bibfield{author}{\bibinfo{person}{Vincent Aleven}, \bibinfo{person}{Jonathan
  Sewall}, \bibinfo{person}{Octav Popescu}, \bibinfo{person}{Michael
  Ringenberg}, \bibinfo{person}{Martin van Velsen}, {and}
  \bibinfo{person}{Sandra Demi}.} \bibinfo{year}{2016}\natexlab{}.
\newblock \showarticletitle{{Embedding Intelligent Tutoring Systems in MOOCs
  and e-Learning Platforms}}. In \bibinfo{booktitle}{\emph{Intelligent Tutoring
  Systems}}, \bibfield{editor}{\bibinfo{person}{Alessandro Micarelli},
  \bibinfo{person}{John Stamper}, {and} \bibinfo{person}{Kitty Panourgia}}
  (Eds.). \bibinfo{publisher}{Springer International Publishing},
  \bibinfo{address}{Cham}, \bibinfo{pages}{409--415}.
\newblock
\showISBNx{978-3-319-39583-8}


\bibitem[\protect\citeauthoryear{Bloom}{Bloom}{1968}]%
        {Bloom1968}
\bibfield{author}{\bibinfo{person}{B Bloom}.} \bibinfo{year}{1968}\natexlab{}.
\newblock \showarticletitle{{Learning for Mastery. Instruction and Curriculum.
  Regional Education Laboratory for the Carolinas and Virginia, Topical Papers
  and Reprints, Number 1.}}
\newblock \bibinfo{journal}{\emph{Evaluation comment}}  \bibinfo{volume}{1}
  (\bibinfo{year}{1968}), \bibinfo{pages}{12}.
\newblock


\bibitem[\protect\citeauthoryear{{Center for Distributed Learning}}{{Center for
  Distributed Learning}}{[n.d.]}]%
        {CenterforDistributedLearning}
\bibfield{author}{\bibinfo{person}{{Center for Distributed Learning}}.}
  \bibinfo{year}{[n.d.]}\natexlab{}.
\newblock \bibinfo{title}{{Obojobo}}.
\newblock
\newblock


\bibitem[\protect\citeauthoryear{Chen, Whitcomb, and Singh}{Chen
  et~al\mbox{.}}{2018}]%
        {Chen2018}
\bibfield{author}{\bibinfo{person}{Zhongzhou Chen}, \bibinfo{person}{Kyle~M.
  Whitcomb}, {and} \bibinfo{person}{Chandralekha Singh}.}
  \bibinfo{year}{2018}\natexlab{}.
\newblock \showarticletitle{Measuring the effectiveness of online
  problem-solving tutorials by multi-level knowledge transfer}. In
  \bibinfo{booktitle}{\emph{Physics Education Research Conference 2018}}
  \emph{(\bibinfo{series}{PER Conference})}. \bibinfo{address}{Washington, DC}.
\newblock


\bibitem[\protect\citeauthoryear{Chen, Xu, Garrido, and Guthrie}{Chen
  et~al\mbox{.}}{2020}]%
        {Chen2020}
\bibfield{author}{\bibinfo{person}{Zhongzhou Chen}, \bibinfo{person}{Mengyu
  Xu}, \bibinfo{person}{Geoffrey Garrido}, {and} \bibinfo{person}{Matthew~W.
  Guthrie}.} \bibinfo{year}{2020}\natexlab{}.
\newblock \showarticletitle{{Relationship between students' online learning
  behavior and course performance: What contextual information matters?}}
\newblock \bibinfo{journal}{\emph{Physical Review Physics Education Research}}
  \bibinfo{volume}{16}, \bibinfo{number}{1} (\bibinfo{date}{jun}
  \bibinfo{year}{2020}), \bibinfo{pages}{010138}.
\newblock
\showISSN{2469-9896}
\urldef\tempurl%
\url{https://doi.org/10.1103/PhysRevPhysEducRes.16.010138}
\showDOI{\tempurl}


\bibitem[\protect\citeauthoryear{Dorn, Hancock, Sarakatsannis, and
  Viruleg}{Dorn et~al\mbox{.}}{2020}]%
        {Dorn2020}
\bibfield{author}{\bibinfo{person}{Emma Dorn}, \bibinfo{person}{Bryan Hancock},
  \bibinfo{person}{Jimmy Sarakatsannis}, {and} \bibinfo{person}{Ellen
  Viruleg}.} \bibinfo{year}{2020}\natexlab{}.
\newblock \bibinfo{title}{{COVID-19 and student learning in the United States:
  The hurt could last a lifetime}}.  (\bibinfo{year}{2020}),
  \bibinfo{numpages}{14}~pages.
\newblock


\bibitem[\protect\citeauthoryear{Eisinga, Heskes, Pelzer, and {Te
  Grotenhuis}}{Eisinga et~al\mbox{.}}{2017}]%
        {Eisinga2017}
\bibfield{author}{\bibinfo{person}{Rob Eisinga}, \bibinfo{person}{Tom Heskes},
  \bibinfo{person}{Ben Pelzer}, {and} \bibinfo{person}{Manfred {Te
  Grotenhuis}}.} \bibinfo{year}{2017}\natexlab{}.
\newblock \showarticletitle{{Exact p-values for pairwise comparison of Friedman
  rank sums, with application to comparing classifiers}}.
\newblock \bibinfo{journal}{\emph{BMC Bioinformatics}} \bibinfo{volume}{18},
  \bibinfo{number}{1} (\bibinfo{year}{2017}), \bibinfo{pages}{1--18}.
\newblock
\showISSN{14712105}
\urldef\tempurl%
\url{https://doi.org/10.1186/s12859-017-1486-2}
\showDOI{\tempurl}


\bibitem[\protect\citeauthoryear{Elliot and Murayama}{Elliot and
  Murayama}{2008}]%
        {Elliot2008}
\bibfield{author}{\bibinfo{person}{Andrew~J Elliot} {and} \bibinfo{person}{Kou
  Murayama}.} \bibinfo{year}{2008}\natexlab{}.
\newblock \showarticletitle{{On the measurement of achievement goals: Critique,
  illustration, and application.}}
\newblock \bibinfo{journal}{\emph{Journal of Educational Psychology}}
  \bibinfo{volume}{100}, \bibinfo{number}{3} (\bibinfo{year}{2008}),
  \bibinfo{pages}{613--628}.
\newblock
\showISSN{1939-2176(Electronic),0022-0663(Print)}
\urldef\tempurl%
\url{https://doi.org/10.1037/0022-0663.100.3.613}
\showDOI{\tempurl}


\bibitem[\protect\citeauthoryear{Felker and Chen}{Felker and Chen}{2020}]%
        {Felker2020}
\bibfield{author}{\bibinfo{person}{Zachary Felker} {and}
  \bibinfo{person}{Zhongzhou Chen}.} \bibinfo{year}{2020}\natexlab{}.
\newblock \showarticletitle{{The impact of extra credit incentives on students'
  work habits when completing online homework assignments}}. In
  \bibinfo{booktitle}{\emph{2020 Physics Education Research Conference
  Proceedings}}. \bibinfo{publisher}{American Association of Physics Teachers
  (AAPT)}, \bibinfo{address}{Virtual Conference}, \bibinfo{pages}{143--148}.
\newblock
\urldef\tempurl%
\url{https://doi.org/10.1119/perc.2020.pr.felker}
\showDOI{\tempurl}


\bibitem[\protect\citeauthoryear{Gonzalez, de~la Rubia, Hincz, Comas-Lopez,
  Subirats, Fort, and Sacha}{Gonzalez et~al\mbox{.}}{2020}]%
        {Gonzalez2020}
\bibfield{author}{\bibinfo{person}{T. Gonzalez}, \bibinfo{person}{M.~A. de~la
  Rubia}, \bibinfo{person}{K.~P. Hincz}, \bibinfo{person}{M. Comas-Lopez},
  \bibinfo{person}{Laia Subirats}, \bibinfo{person}{Santi Fort}, {and}
  \bibinfo{person}{G.~M. Sacha}.} \bibinfo{year}{2020}\natexlab{}.
\newblock \showarticletitle{{Influence of COVID-19 confinement on students'
  performance in higher education}}.
\newblock \bibinfo{journal}{\emph{PloS one}} \bibinfo{volume}{15},
  \bibinfo{number}{10} (\bibinfo{year}{2020}), \bibinfo{pages}{e0239490}.
\newblock
\showISBNx{1111111111}
\showISSN{19326203}
\urldef\tempurl%
\url{https://doi.org/10.1371/journal.pone.0239490}
\showDOI{\tempurl}


\bibitem[\protect\citeauthoryear{Guthrie, Zhang, and Chen}{Guthrie
  et~al\mbox{.}}{2020}]%
        {Guthrie2020}
\bibfield{author}{\bibinfo{person}{Matthew~W. Guthrie}, \bibinfo{person}{Tom
  Zhang}, {and} \bibinfo{person}{Zhongzhou Chen}.}
  \bibinfo{year}{2020}\natexlab{}.
\newblock \showarticletitle{{A tale of two guessing strategies: interpreting
  the time students spend solving problems through online log data}}. In
  \bibinfo{booktitle}{\emph{Physics Education Research Conference
  Proceedings}}. \bibinfo{publisher}{American Association of Physics Teachers
  (AAPT)}, \bibinfo{address}{Virtual Conference}, \bibinfo{pages}{185--190}.
\newblock
\urldef\tempurl%
\url{https://doi.org/10.1119/perc.2020.pr.guthrie}
\showDOI{\tempurl}


\bibitem[\protect\citeauthoryear{Gutmann, Gladding, Lundsgaard, and
  Stelzer}{Gutmann et~al\mbox{.}}{[n.d.]}]%
        {Gutmann}
\bibfield{author}{\bibinfo{person}{Brianne Gutmann}, \bibinfo{person}{Gary~E.
  Gladding}, \bibinfo{person}{Morten Lundsgaard}, {and}
  \bibinfo{person}{Timothy Stelzer}.} \bibinfo{year}{[n.d.]}\natexlab{}.
\newblock \showarticletitle{{Mastery-style homework exercises in introductory
  physics courses: Implementation matters}}.
\newblock \bibinfo{journal}{\emph{Physical Review Physics Education Research}}
  (\bibinfo{year}{[n.\,d.]}).
\newblock


\bibitem[\protect\citeauthoryear{Li, Baker, and Warschauer}{Li
  et~al\mbox{.}}{2020}]%
        {Li2020}
\bibfield{author}{\bibinfo{person}{Qiujie Li}, \bibinfo{person}{Rachel Baker},
  {and} \bibinfo{person}{Mark Warschauer}.} \bibinfo{year}{2020}\natexlab{}.
\newblock \showarticletitle{{Using clickstream data to measure, understand, and
  support self-regulated learning in online courses}}.
\newblock \bibinfo{journal}{\emph{Internet and Higher Education}}
  \bibinfo{volume}{45} (\bibinfo{date}{apr} \bibinfo{year}{2020}),
  \bibinfo{pages}{100727}.
\newblock
\showISSN{10967516}
\urldef\tempurl%
\url{https://doi.org/10.1016/j.iheduc.2020.100727}
\showDOI{\tempurl}


\bibitem[\protect\citeauthoryear{Maldonado-Mahauad,
  P{\'{e}}rez-Sanagust{\'{i}}n, Kizilcec, Morales, and
  Munoz-Gama}{Maldonado-Mahauad et~al\mbox{.}}{2018}]%
        {Maldonado-Mahauad2018}
\bibfield{author}{\bibinfo{person}{Jorge Maldonado-Mahauad},
  \bibinfo{person}{Mar P{\'{e}}rez-Sanagust{\'{i}}n},
  \bibinfo{person}{Ren{\'{e}}~F. Kizilcec}, \bibinfo{person}{Nicol{\'{a}}s
  Morales}, {and} \bibinfo{person}{Jorge Munoz-Gama}.}
  \bibinfo{year}{2018}\natexlab{}.
\newblock \showarticletitle{{Mining theory-based patterns from Big data:
  Identifying self-regulated learning strategies in Massive Open Online
  Courses}}.
\newblock \bibinfo{journal}{\emph{Computers in Human Behavior}}
  \bibinfo{volume}{80} (\bibinfo{year}{2018}), \bibinfo{pages}{179--196}.
\newblock
\showISSN{07475632}
\urldef\tempurl%
\url{https://doi.org/10.1016/j.chb.2017.11.011}
\showDOI{\tempurl}


\bibitem[\protect\citeauthoryear{Pohlert}{Pohlert}{2020}]%
        {PMCMRplus}
\bibfield{author}{\bibinfo{person}{Thorsten Pohlert}.}
  \bibinfo{year}{2020}\natexlab{}.
\newblock \bibinfo{booktitle}{\emph{PMCMRplus: Calculate Pairwise Multiple
  Comparisons of Mean Rank Sums Extended}}.
\newblock
\urldef\tempurl%
\url{https://CRAN.R-project.org/package=PMCMRplus}
\showURL{%
\tempurl}
\newblock
\shownote{R package version 1.6.1.}


\bibitem[\protect\citeauthoryear{{R Core Team}}{{R Core Team}}{2019}]%
        {RCoreTeam2019}
\bibfield{author}{\bibinfo{person}{{R Core Team}}.}
  \bibinfo{year}{2019}\natexlab{}.
\newblock \bibinfo{title}{{R: A Language and Environment for Statistical
  Computing}}.
\newblock
\newblock


\bibitem[\protect\citeauthoryear{Schwartz and Bransford}{Schwartz and
  Bransford}{2005}]%
        {Schwartz2005}
\bibfield{author}{\bibinfo{person}{Daniel~L. Schwartz} {and}
  \bibinfo{person}{John~D. Bransford}.} \bibinfo{year}{2005}\natexlab{}.
\newblock \showarticletitle{{Efficiency and Innovation in Transfer}}.
\newblock In \bibinfo{booktitle}{\emph{Transfer of Learning from a Modern
  Multidisciplinary Perspective (Current Perspectives on Cognition, Learning
  and Instruction)}}, \bibfield{editor}{\bibinfo{person}{Jose~P. Mestre}}
  (Ed.). \bibinfo{publisher}{IAP - Informaiton Age Publishing Inc.},
  \bibinfo{address}{Charlotte}, \bibinfo{pages}{1--51}.
\newblock
\showISBNx{1593111649}


\bibitem[\protect\citeauthoryear{Taub, Azevedo, Bradbury, Millar, and
  Lester}{Taub et~al\mbox{.}}{2018}]%
        {Taub2018}
\bibfield{author}{\bibinfo{person}{Michelle Taub}, \bibinfo{person}{Roger
  Azevedo}, \bibinfo{person}{Amanda~E. Bradbury}, \bibinfo{person}{Garrett~C.
  Millar}, {and} \bibinfo{person}{James Lester}.}
  \bibinfo{year}{2018}\natexlab{}.
\newblock \showarticletitle{{Using sequence mining to reveal the efficiency in
  scientific reasoning during STEM learning with a game-based learning
  environment}}.
\newblock \bibinfo{journal}{\emph{Learning and Instruction}}
  \bibinfo{volume}{54} (\bibinfo{year}{2018}), \bibinfo{pages}{93--103}.
\newblock
\showISSN{09594752}
\urldef\tempurl%
\url{https://doi.org/10.1016/j.learninstruc.2017.08.005}
\showDOI{\tempurl}


\bibitem[\protect\citeauthoryear{{U.S Department of Education} and {Office of
  Educational Technology}}{{U.S Department of Education} and {Office of
  Educational Technology}}{2017}]%
        {U.SDepartmentofEducation2017}
\bibfield{author}{\bibinfo{person}{{U.S Department of Education}} {and}
  \bibinfo{person}{{Office of Educational Technology}}.}
  \bibinfo{year}{2017}\natexlab{}.
\newblock \bibinfo{title}{{Reimagining the Role of Technology in Higher
  Education}}.  (\bibinfo{year}{2017}).
\newblock


\bibitem[\protect\citeauthoryear{Wickham, Averick, Bryan, Chang, McGowan,
  François, Grolemund, Hayes, Henry, Hester, Kuhn, Pedersen, Miller, Bache,
  Müller, Ooms, Robinson, Seidel, Spinu, Takahashi, Vaughan, Wilke, Woo, and
  Yutani}{Wickham et~al\mbox{.}}{2019}]%
        {tidyverse}
\bibfield{author}{\bibinfo{person}{Hadley Wickham}, \bibinfo{person}{Mara
  Averick}, \bibinfo{person}{Jennifer Bryan}, \bibinfo{person}{Winston Chang},
  \bibinfo{person}{Lucy~D'Agostino McGowan}, \bibinfo{person}{Romain
  François}, \bibinfo{person}{Garrett Grolemund}, \bibinfo{person}{Alex
  Hayes}, \bibinfo{person}{Lionel Henry}, \bibinfo{person}{Jim Hester},
  \bibinfo{person}{Max Kuhn}, \bibinfo{person}{Thomas~Lin Pedersen},
  \bibinfo{person}{Evan Miller}, \bibinfo{person}{Stephan~Milton Bache},
  \bibinfo{person}{Kirill Müller}, \bibinfo{person}{Jeroen Ooms},
  \bibinfo{person}{David Robinson}, \bibinfo{person}{Dana~Paige Seidel},
  \bibinfo{person}{Vitalie Spinu}, \bibinfo{person}{Kohske Takahashi},
  \bibinfo{person}{Davis Vaughan}, \bibinfo{person}{Claus Wilke},
  \bibinfo{person}{Kara Woo}, {and} \bibinfo{person}{Hiroaki Yutani}.}
  \bibinfo{year}{2019}\natexlab{}.
\newblock \showarticletitle{Welcome to the {tidyverse}}.
\newblock \bibinfo{journal}{\emph{Journal of Open Source Software}}
  \bibinfo{volume}{4}, \bibinfo{number}{43} (\bibinfo{year}{2019}),
  \bibinfo{pages}{1686}.
\newblock
\urldef\tempurl%
\url{https://doi.org/10.21105/joss.01686}
\showDOI{\tempurl}


\bibitem[\protect\citeauthoryear{Wilcox and Vignal}{Wilcox and Vignal}{2020}]%
        {Wilcox2020}
\bibfield{author}{\bibinfo{person}{Bethany~R. Wilcox} {and}
  \bibinfo{person}{Michael Vignal}.} \bibinfo{year}{2020}\natexlab{}.
\newblock \showarticletitle{{Understanding the student experience with
  emergency remote teaching}}. In \bibinfo{booktitle}{\emph{2020 Physics
  Education Research Conference Proceedings}}. \bibinfo{publisher}{American
  Association of Physics Teachers (AAPT)}, \bibinfo{address}{Virtual
  Conference}, \bibinfo{pages}{581--586}.
\newblock
\urldef\tempurl%
\url{https://doi.org/10.1119/perc.2020.pr.wilcox}
\showDOI{\tempurl}


\bibitem[\protect\citeauthoryear{Zimmerman}{Zimmerman}{2013}]%
        {Zimmerman2013}
\bibfield{author}{\bibinfo{person}{Barry~J. Zimmerman}.}
  \bibinfo{year}{2013}\natexlab{}.
\newblock \showarticletitle{{From Cognitive Modeling to Self-Regulation: A
  Social Cognitive Career Path}}.
\newblock \bibinfo{journal}{\emph{Educational Psychologist}}
  \bibinfo{volume}{48}, \bibinfo{number}{3} (\bibinfo{year}{2013}),
  \bibinfo{pages}{135--147}.
\newblock
\showISSN{00461520}
\urldef\tempurl%
\url{https://doi.org/10.1080/00461520.2013.794676}
\showDOI{\tempurl}


\bibitem[\protect\citeauthoryear{Zimmerman}{Zimmerman}{2015}]%
        {Zimmerman2015}
\bibfield{author}{\bibinfo{person}{Barry~J. Zimmerman}.}
  \bibinfo{year}{2015}\natexlab{}.
\newblock \bibinfo{booktitle}{\emph{{Self-Regulated Learning: Theories,
  Measures, and Outcomes}} (\bibinfo{edition}{second edi} ed.)}.
  Vol.~\bibinfo{volume}{21}.
\newblock \bibinfo{publisher}{Elsevier}. 541--546 pages.
\newblock
\showISBNx{9780080970868}
\urldef\tempurl%
\url{https://doi.org/10.1016/B978-0-08-097086-8.26060-1}
\showDOI{\tempurl}


\end{thebibliography}


\end{document}